\documentclass[aps,prd,twocolumn,showkeys,amsmath,amssymb]{revtex4-2}
\usepackage{graphicx}
\usepackage{epstopdf}
\usepackage{multirow}
\usepackage{makecell}
\usepackage{subfigure}
\usepackage{extarrows}
\usepackage{feynmf}
\usepackage{enumitem}
\usepackage{color}
\usepackage[colorlinks,citecolor=blue,anchorcolor=red,menucolor=red,linkcolor=red,filecolor=red,runcolor=red,urlcolor=blue]{hyperref}

\begin{document}

\title{A toy model for hadronic bonds in exotic hadrons}

\author{Hua-Xing Chen}
\email{hxchen@seu.edu.cn}
\affiliation{School of Physics, Southeast University, Nanjing 210094, China}

\begin{abstract}
Three hadronic-bond mechanisms were proposed in our previous studies~\cite{Chen:2021xlu,Chen:2025jgi} to describe near-threshold exotic hadrons in terms of hadronic molecular configurations: the covalent bond arises from the sharing and exchange of valence light quarks between hadronic clusters, the creation bond originates from the activation of coherent sea $q\bar{q}$ degrees of freedom followed by the formation of effective covalent correlations, and the annihilation bond arises from valence $q\bar{q}$ annihilation and the associated mixing with a nearby compact state. A simple toy model was previously constructed only for the covalent bond, and in the present study we extend this model to incorporate the creation and annihilation bonds within a unified framework. The resulting framework organizes covalent-bond configurations as tending to lie below threshold and creation-bond configurations as tending to appear above threshold, while the annihilation bond can shift a pre-existing molecular configuration either downward or upward through mixing with a nearby compact state. An effective activation energy $E_{\rm{act}}$ is introduced for the coherent sea four-quark component involved in the creation bond, with the $Z_c(3900)$ providing the benchmark value $E_{\rm{act}}\sim17~{\rm{MeV}}$. Intriguingly, this value is numerically close to the reported mass of the $X_{17}$ particle, which we previously interpreted as a low-energy bare four-quark structure~\cite{Chen:2020arr}. This numerical proximity may therefore reflect common low-energy (four-quark) dynamics, although the microscopic connection remains to be established. For the isoscalar $D^*\bar{D}^*B^*\bar{B}^*$ configuration involving the $Z_c(4020)$ and $Z_b(10650)$ creation-bond subsystems linked by two additional covalent bonds, we estimate a binding energy of approximately $36~{\rm{MeV}}$ below its four-meson threshold and propose searches in the $4\mu$ and $4\mu2\pi$ final states.
\end{abstract}

\date{\today}
\maketitle

\section{Introduction}
\label{sec:intro}

The study of exotic hadrons has become an important subject in hadron spectroscopy following the observation of a large number of states that cannot be straightforwardly accommodated within the conventional quark model~\cite{pdg}. Many of these structures appear close to two-hadron thresholds and are therefore naturally associated with hadronic molecular configurations~\cite{Weinberg:1965zz,Voloshin:1976ap,DeRujula:1976zlg,Tornqvist:1993ng,Voloshin:2003nt,Close:2003sg,Wong:2003xk,Braaten:2003he,Swanson:2003tb,Tornqvist:2004qy,Wu:2010jy,Wang:2011rga,Yang:2011wz,Karliner:2015ina,Chen:2015loa,Chen:2015moa,Liu:2019tjn}, while compact multiquark components may also play important roles~\cite{Maiani:2004vq,Maiani:2014aja,Hogaasen:2005jv,Ebert:2005nc,Barnea:2006sd}. The rapidly growing number of such candidates motivates the search for simple organizing principles that can correlate their properties across different flavor and spin sectors~\cite{Chen:2016qju,Liu:2019zoy,Chen:2022asf,Luo:2025sns,Hosaka:2016pey,Richard:2016eis,Lebed:2016hpi,Esposito:2016noz,Ali:2017jda,Guo:2017jvc,Olsen:2017bmm,Karliner:2017qhf,Guo:2019twa,Brambilla:2019esw,Yang:2020atz,Fang:2021wes,Jin:2021vct,Meng:2022ozq,Liu:2024uxn,Wang:2025dur,Bai:2026atm,Dai:2026fkg}.

\begin{figure*}[t]
\centering
\includegraphics[width=0.9\textwidth]{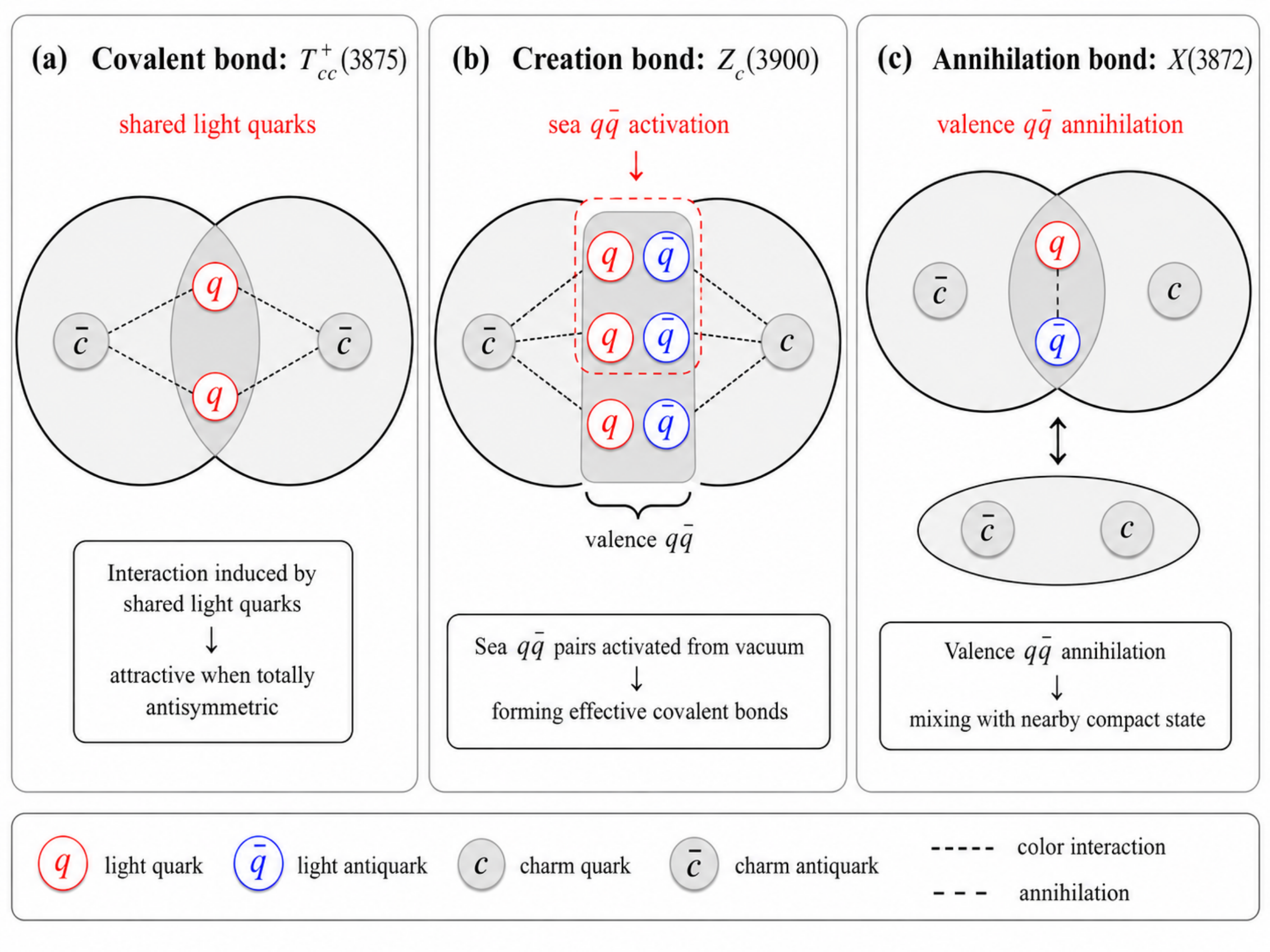}
\caption{Schematic illustration of the three hadronic-bond mechanisms considered in this work. (a) Covalent bond arising from the sharing and exchange of valence light quarks or antiquarks, with \(T_{cc}(3875)\) as a representative example. (b) Creation bond arising from coherent sea \(q\bar q\) activation followed by the formation of effective covalent bonds, with \(Z_c(3900)\) as a representative example. (c) Annihilation bond arising from valence \(q\bar q\) annihilation and mixing with a nearby compact state, with \(X(3872)\) as a representative example.}
\label{fig:bonds}
\end{figure*}

Our basic assumption is that the residual interactions among color-singlet hadronic clusters are governed primarily by the light-quark degrees of freedom and can be organized according to the underlying quark-line processes, as illustrated in Fig.~\ref{fig:bonds}. Motivated by our analyses based on the QCD sum rule method~\cite{Shifman:1978bx,Reinders:1984sr,Colangelo:2000dp,Narison:2002pw,Albuquerque:2018jkn,Wang:2025sic}, we successively proposed three types of hadronic bonds in Refs.~\cite{Chen:2021xlu,Chen:2025jgi} to describe these residual interactions, with \(q=u,d\) denoting a light quark:
\begin{itemize}

\item \textit{Covalent bond.}
It arises from the direct sharing and exchange of valence light quarks between color-singlet hadronic clusters~\cite{Oka:1981ri,Oka:1981rj,Hoodbhoy:1986fn,Modarres:1988sb,Wang:1992wi,Rosina:1994uqr,Modarres:2006ry}. In analogy with chemical covalent bonding, the combined color, flavor, spin, and orbital wave function of the shared identical quarks must be overall antisymmetric, as required by Fermi statistics.

\item \textit{Creation bond.}
It arises when sea \(q\bar q\) degrees of freedom participate in the interhadronic interaction and generate effective covalent correlations in cases where the direct sharing of valence light quarks is not available. This mechanism requires an activation energy, so that the resulting structure may appear above the corresponding threshold.

\item \textit{Annihilation bond.}
It originates from valence \(q\bar q\) annihilation and acts on a two-hadron configuration already correlated through the covalent or creation bond. In the absence of a nearby compact state with the same quantum numbers, annihilation may predominantly broaden or dissolve the molecular component. When such a compact state is available, annihilation can induce molecular--compact mixing, producing two shifted eigenstates---of which only one may be experimentally identifiable---with the molecular-dominated state shifted either downward or upward.

\end{itemize}

In Ref.~\cite{Chen:2021xlu}, we constructed a simple toy model for the covalent bond, in which the binding energy is approximated as a sum of bond contributions characterized by five phenomenological parameters. This construction provided a quantitative realization of the covalent-bond picture, but the creation and annihilation bonds introduced subsequently were not incorporated into the same toy model. In the present work, we extend this model to incorporate both mechanisms within a unified framework. Only two additional phenomenological parameters are required: an effective activation energy \(E_{\rm act}\) for the creation bond and a signed spectral shift \(\Delta\) associated with annihilation-induced molecular--compact mixing.

Taking the \(Z_c(3900)\) as a benchmark gives \(E_{\rm act}\sim17~{\rm MeV}\). Intriguingly, this value is numerically close to the reported mass of the \(X_{17}\) particle, which we previously interpreted as a low-energy bare four-quark structure~\cite{Chen:2020arr}. The \(X_{17}\) particle was first reported by Krasznahorkay et al. at ATOMKI~\cite{Krasznahorkay:2015iga,Krasznahorkay:2017qfd,Krasznahorkay:2019lyl,Krasznahorkay:2022pxs}, with similar results subsequently reported by VINATOM~\cite{Anh:2024req}. Although various explanations have been proposed~\cite{Feng:2016jff,Feng:2016ysn,Gu:2016ege,Kahn:2016vjr,Fayet:2016nyc,Dror:2017nsg,Liang:2016ffe,Jia:2016uxs,Ellwanger:2016wfe,Kozaczuk:2016nma,Dror:2017ehi,Hati:2020fzp,Zhang:2017zap,Veselsky:2020ewb,Bauer:2017ris,Alves:2017avw,Kirpichnikov:2020tcf,Dusaev:2020gxi,Wong:2020hjc,Wong:2021blz}, its nature remains unsettled~\cite{Barducci:2022lqd,Alves:2023ree,Cirelli:2024ssz,Serao:2026zeu}.

The remainder of this paper is organized as follows. In Sec.~\ref{sec:framework}, we formulate the hadronic-bond framework and construct the toy-model descriptions of the covalent, creation, and annihilation bonds. In Sec.~\ref{sec:analysis}, we analyze representative hadronic systems, discuss several predictions of the model, and explore extensions to multihadron and kaon-related systems. Finally, we summarize our main results and conclusions in Sec.~\ref{sec:summary}.

\begin{table*}[t]
\centering
\renewcommand{\arraystretch}{1.6}
\caption{
Generalized binding energies of representative systems within the hadronic-bond framework. The generalized binding energy is defined as \(B=M_{\rm threshold}-M\), such that \(B>0\) (\(B<0\)) places the system below (above) the relevant threshold. For hadronic systems, the thresholds are evaluated using the isospin-averaged central masses of the constituent hadrons quoted by the Particle Data Group, and the corresponding experimental values, including those in the last column, are approximate values derived from these central masses. For nuclei and hypernuclei, \(B_{\rm exp}\) is the total binding energy relative to the corresponding physical free \(p\), \(n\), and \(\Lambda\) constituents, rather than a hyperon separation energy. The calculated values are benchmark estimates within the present phenomenological framework. The third column lists the adopted hadronic configurations together with their quantum numbers \(I(J^P)\), or \(I^G(J^{PC})\) where applicable. For exotic-hadron candidates, the listed hadronic configurations represent the molecular-channel assignments adopted in the present framework and are not necessarily experimentally established. Some candidates listed in the last column are tentative and are included for phenomenological comparison. The annihilation-induced contribution is treated as a signed quantity with benchmark magnitude \(|\Delta|\sim15~{\rm MeV}\): \(\Delta\sim+15~{\rm MeV}\) is used for \(X(3872)\), whereas \(\Delta\sim-15~{\rm MeV}\) is used for \(X(3960)\). A positive (negative) \(\Delta\) increases (decreases) the calculated binding energy. For \(Z_c(3900)\) and \(X(3872)\), \(D\bar D^*\) denotes the appropriate coherent combination of the \(D\bar D^*\) and \(D^*\bar D\) channels, and analogous shorthand notation is used for related mixed channels where applicable. The values \(B_{\rm exp}\simeq4~{\rm MeV}\) for \(X(3872)\) and \(B_{\rm exp}\simeq1~{\rm MeV}\) for \(T_{cc}(3875)\) refer to isospin-averaged thresholds; relative to the physical \(D^0\bar D^{*0}\) and \(D^0D^{*+}\) thresholds, the corresponding binding energies are of order \(0.1~{\rm MeV}\) and \(0.36~{\rm MeV}\), respectively.
}
\label{tab:results}
\begin{tabular}{c c c c c c c c c c c l}
\hline
Bond type
& System
& Configuration
& \(B_{\rm exp}\)
& \(\Delta\)
& \(N_S\)
& \(N_W\)
& \(N_\Lambda\)
& \(N_R\)
& \(N_h\)
& \(B_{\rm calc}\)
& Related systems

\\ \hline

\multirow{12}{*}{Covalent}
& \(T_{cc}(3875)\)
& \(DD^*[0(1^+)]\)
& \(1~{\rm MeV}\)
& --
& 1
& 0
& 0
& 1
& 2
& \(1~{\rm MeV}\)
& 
\(d:pn~[2~{\rm MeV}]\)

\\
& \(P_c(4312)\)
& \(\bar D \Sigma_c[1/2(1/2^-)]\)
& \(9~{\rm MeV}\)
& --
& 1
& 0
& 0
& 0
& 2
& \(15~{\rm MeV}\)
& 

\makecell[l]{
\\[-0.2cm]
\(P_c(4380):\bar D\Sigma_c^*[5~{\rm MeV}]\)
\\
\(P_c(4440):\bar D^* \Sigma_c[22~{\rm MeV}]\)
\\
\(P_c(4457):\bar D^* \Sigma_c[5~{\rm MeV}]\)
\\
\(P_{cs}(4338):\bar D \Xi_c[-2~{\rm MeV}]\)
\\
\(P_{cs}(4459):\bar D^* \Xi_c[19~{\rm MeV}]\)
\\
\(T_{\bar c \bar s 0}^*(2870):\bar D^* K^*[19~{\rm MeV}]\)
\\[0.1cm]
}

\\
& \({}^{3}{\rm He}\)
& \(ppn[1/2(1/2^+)]\)
& \(8~{\rm MeV}\)
& --
& 2
& 1
& 0
& 3
& 3
& \(9~{\rm MeV}\)
& 
\(^{3}{\rm H}:pnn~[8~{\rm MeV}]\)

\\
& \(^{4}{\rm He}\)
& \(ppnn[0(0^+)]\)
& \(28~{\rm MeV}\)
& --
& 4
& 2
& 0
& 6
& 4
& \(26~{\rm MeV}\)
& 

\\
& \({}^{3}_{\Lambda}{\rm H}\)
& \(pn\Lambda[0(1/2^+)]\)
& \(2~{\rm MeV}\)
& --
& 1
& 0
& 1
& 2
& 3
& \(1~{\rm MeV}\)
& 

\\
& \({}^{4}_{\Lambda}{\rm He}\)
& \(ppn\Lambda[1/2(0^+)]\)
& \(10~{\rm MeV}\)
& --
& 2
& 1
& 2
& 5
& 4
& \(13~{\rm MeV}\)
& 
\({}^{4}_{\Lambda}{\rm H}:pnn\Lambda~[11~{\rm MeV}]\)

\\
& \({}^{5}_{\Lambda}{\rm He}\)
& \(ppnn\Lambda[0(1/2^+)]\)
& \(31~{\rm MeV}\)
& --
& 4
& 2
& 2
& 8
& 5
& \(30~{\rm MeV}\)
& 

\\
& \({}^{6}_{\Lambda\Lambda}{\rm He}\)
& \(ppnn\Lambda\Lambda[0(0^+)]\)
& \(35~{\rm MeV}\)
& --
& 4
& 2
& 4
& 10
& 6
& \(34~{\rm MeV}\)
& 

\\ \hline

\multirow{3}{*}{Creation}

& \(Z_c(3900)\)
& \(D\bar D^*[1^+(1^{+-})]\)
& \(-11~{\rm MeV}\)
& --
& 4
& 2
& 0
& 0
& 2
& \(-11~{\rm MeV}\)
& 

\makecell[l]{
\\[-0.2cm]
\(Z_c(4020):D^*\bar D^*[-7~{\rm MeV}]\)
\\
\(Z_b(10610):B\bar B^*[-3~{\rm MeV}]\)
\\
\(Z_b(10650):B^*\bar B^*[-3~{\rm MeV}]\)
\\
\(T_{c \bar s 0}^a(2900):D^* K^*[-6~{\rm MeV}]\)
\\[0.1cm]
}

\\
& \(Z_{cs}(3985)\)
& \(D_s^*\bar D[1/2(1^+)]\)
& \(-4~{\rm MeV}\)
& --
& 3
& 1
& 1
& 0
& 2
& \(-13~{\rm MeV}\)
& 
\(Z_{cs}(4000):D_s^*\bar D[-25~{\rm MeV}]\)

\\ \hline
\rule{0pt}{0.55cm}

\multirow{2.5}{*}{
\makecell[c]{Creation \&
\\[0.2cm]
Annihilation
}}

& \(X(3872)\)
& \(D\bar D^*[0^+(1^{++})]\)
& \(4~{\rm MeV}\)
& \(+15~{\rm MeV}\)
& 4
& 2
& 0
& 0
& 2
& \(4~{\rm MeV}\)
& 
\(\Lambda_c(2940):D^*N[8~{\rm MeV}]\)

\\[0.2cm]
& \(X(3960)\)
& \(D_s \bar D_s[0^+(0^{++})]\)
& \(-19~{\rm MeV}\)
& \(-15~{\rm MeV}\)
& 2
& 0
& 2
& 0
& 2
& \(-31~{\rm MeV}\)
& 

\\[0.1cm] \hline
\end{tabular}
\end{table*}

\section{Hadronic-Bond Framework}
\label{sec:framework}

In this section we formulate a minimal hadronic-bond framework for organizing near-threshold exotic-hadron candidates. As illustrated schematically in Fig.~\ref{fig:bonds}, we classify the effective correlations among color-singlet hadrons into three types of hadronic bonds: the covalent bond, the creation bond, and the annihilation bond. These three mechanisms emphasize different roles played by light quarks and antiquarks in forming exotic structures. The covalent bond arises from the sharing and exchange of valence light quarks or antiquarks between two hadronic clusters, the creation bond originates from the activation of coherent sea \(q\bar q\) pairs followed by the formation of effective covalent bonds, and the annihilation bond arises from valence \(q\bar q\) annihilation and the associated mixing with a nearby compact state.

Together, these three mechanisms constitute the basic ingredients of the present phenomenological framework. In the following subsections, we introduce them separately, using \(T_{cc}(3875)\), \(Z_c(3900)\), and \(X(3872)\) as representative examples of the covalent, creation, and annihilation bonds, respectively. Note that these three bonds do not represent fundamental interactions beyond QCD. Instead, they provide effective descriptions of the residual correlations among color-singlet hadrons generated by different light-quark processes, including light-quark exchange, sea \(q\bar q\) activation, and valence \(q\bar q\) annihilation.

Throughout this work, we characterize near-threshold hadronic structures by the generalized binding energy
\begin{equation}
B\equiv M_{\rm threshold}-M,
\label{eq:Bdef}
\end{equation}
where \(M_{\rm threshold}\) denotes the sum of the masses of the noninteracting hadrons defining the corresponding threshold, and \(M\) is the mass of the state under consideration. A positive value of \(B\) indicates a state below its corresponding threshold, while a negative value corresponds to an above-threshold structure. For above-threshold resonances, this quantity is interpreted as a generalized threshold offset rather than a conventional binding energy. This definition can be applied uniformly to systems composed of mesons, baryons, or more general hadronic clusters. Within the present framework, the three hadronic-bond mechanisms contribute differently to the generalized binding energy.

\subsection{Covalent Bond}
\label{subsec:covalent}

The covalent hadronic bond was first proposed in Ref.~\cite{Chen:2021xlu} as an effective interaction mechanism induced by shared light quarks or antiquarks between two hadronic clusters. The basic idea is inspired by chemical covalent bonding, in which two atoms form a stable molecule through the sharing of electrons. In the hadronic analogue, the shared degrees of freedom are light quarks or antiquarks, whose rearrangement and exchange between neighboring hadrons can generate an effective residual attraction. By analogy with chemical covalent bonds, in which the shared electrons obey the antisymmetry requirement imposed by the Pauli principle, we proposed that the light-quark-exchange interaction is attractive when the total wave function of the shared light-quark system is overall antisymmetric. This provides a simple criterion for identifying favorable covalent-bond configurations and estimating their binding tendencies.

A natural prototype for the covalent hadronic bond is the doubly charmed tetraquark candidate \(T_{cc}(3875)\), commonly interpreted as a \(DD^*\) molecular candidate with \(I(J^P)=0(1^+)\). For convenience, we discuss its charge-conjugated counterpart, \(\bar D\bar D^*\), which has the same binding dynamics. As illustrated in Fig.~\ref{fig:bonds}(a), the two heavy-meson clusters can exchange and rearrange their light-quark degrees of freedom, thereby forming an attractive covalent configuration. To illustrate the underlying symmetry structure, let us consider the exchange of the light quarks \(q_1\) and \(q_3\) between the two hadronic clusters, \(\bar D[q_1\bar c_2]\) and \(\bar D^*[q_3\bar c_4]\). The two exchanged light quarks \(q_1\) and \(q_3\) are taken to carry the same color, so their color wave function is symmetric under the exchange \(q_1\leftrightarrow q_3\). Their relative orbital wave function is assumed to be an \(S\)-wave and is therefore also symmetric. Their spin wave function can be either symmetric or antisymmetric, since exchanging the light degrees of freedom interchanges the \(\bar D\bar D^*\) and \(\bar D^*\bar D\) configurations. In the isospin limit, the flavor wave function of the two light quarks can likewise be either symmetric or antisymmetric.

Imposing the Pauli principle, two overall-antisymmetric configurations of the exchanged light quarks can be constructed. The first configuration is
\begin{equation}\nonumber
\begin{array}{cccccc}
\hline
\textit{strong}                 & {\rm color} & {\rm flavor} & {\rm spin} & {\rm orbital} & {\rm total}
\\ \hline
q_1\leftrightarrow q_3          & {\bf S}     & {\bf A}      & {\bf S}    & {\bf S}       & {\bf A}
\\ \hline
\end{array}
\end{equation}
while the second configuration is
\begin{equation}\nonumber
\begin{array}{cccccc}
\hline
\textit{weak}                   & {\rm color} & {\rm flavor} & {\rm spin} & {\rm orbital} & {\rm total}
\\ \hline
q_1\leftrightarrow q_3          & {\bf S}     & {\bf S}      & {\bf A}    & {\bf S}       & {\bf A}
\\ \hline
\end{array}
\end{equation}
Here, \({\bf S}\) and \({\bf A}\) denote wave functions that are symmetric and antisymmetric, respectively. As discussed in Ref.~\cite{Chen:2021xlu}, the first configuration with \(I(J^P)=0(1^+)\) generates a stronger attraction than the second configuration with \(I(J^P)=1(0^+)\). Accordingly, we refer to them as the strong and weak covalent bonds, respectively.

By contrast, the exchange of the two identical heavy antiquarks gives rise to a repulsive contribution:
\begin{equation}\nonumber
\begin{array}{cccccc}
\hline
\textit{repulsive}              & {\rm color} & {\rm flavor} & {\rm spin} & {\rm orbital} & {\rm total}
\\ \hline
\bar c_2\leftrightarrow\bar c_4 & {\bf S}     & {\bf S}      & {\bf S}    & {\bf S}       & {\bf S}
\\ \hline
\end{array}
\end{equation}
The corresponding total wave function is symmetric under the exchange \(\bar c_2\leftrightarrow\bar c_4\) and therefore does not satisfy the antisymmetry requirement imposed by the Pauli principle. Within the present bond-counting scheme, this Pauli-forbidden exchange pattern is represented by an effective repulsive contribution. The \(T_{cc}(3875)\) state is therefore described as a \(\bar D\bar D^*\) covalent molecule with \(I(J^P)=0(1^+)\), containing one strong covalent bond and one repulsive contribution from heavy-antiquark exchange.

The covalent-bond mechanism can be extended to configurations involving strange-quark degrees of freedom. By replacing one of the light \(q=(u,d)\) quarks with a strange quark \(s\), one obtains the strange counterparts of the strong and weak covalent bonds introduced above. In an approximate flavor-\(SU(3)\) classification, the strange analogue of the strong covalent bond, with \(I(J^P)=1/2(1^+)\), is characterized by
\begin{equation}\nonumber
\begin{array}{cccccc}
\hline
\textit{strange}                & {\rm color} & {\rm flavor} & {\rm spin} & {\rm orbital} & {\rm total}
\\ \hline
q_1\leftrightarrow s_3          & {\bf S}     & {\bf A}      & {\bf S}    & {\bf S}       & {\bf A}
\\ \hline
\end{array}
\end{equation}
where the flavor-antisymmetric \(qs\) combination represents the strange analogue of the strong covalent configuration in the approximate flavor-\(SU(3)\) limit. The strange analogue of the weak covalent bond, with \(I(J^P)=1/2(0^+)\), is not considered in the present study because no well-established experimental candidate has yet been identified for this configuration. For simplicity, we refer to the strange analogue of the strong covalent bond as the strange bond in the following.

Within this phenomenological framework, the generalized binding energy of a system described by the covalent-bond mechanism is parameterized as
\begin{equation}
B_{\rm cov}
=
N_S S
+
N_W W
+
N_\Lambda \Lambda
-
N_R R
-
N_h\epsilon ,
\label{eq:Bcov}
\end{equation}
where \(S\), \(W\), and \(\Lambda\) denote the effective strengths of the strong, weak, and strange covalent bonds defined above, respectively. Here, \(\Lambda\) denotes the strange-bond strength rather than the \(\Lambda\) hyperon. These phenomenological parameters encode the net effects of the color, flavor, spin, and orbital symmetries of the corresponding shared light-quark configurations. The coefficients \(N_S\), \(N_W\), and \(N_\Lambda\) count the strong, weak, and strange covalent bonds in a given hadronic configuration, respectively. Similarly, \(N_R\) counts the repulsive exchange contributions, while \(N_h\) denotes the number of constituent hadronic clusters. The parameter \(R\) represents the effective strength of the repulsive contribution associated with Pauli-disfavored exchange patterns, whereas \(\epsilon\) represents an effective residual energy cost per hadronic cluster that accounts for kinetic and finite-size effects.

Importantly, the parameters in Eq.~\eqref{eq:Bcov} are not fitted to exotic-hadron masses. Instead, they were estimated phenomenologically in Ref.~\cite{Chen:2021xlu} from light nuclear and hypernuclear systems~\cite{Wang:2021xhn}, where the covalent-bond picture was applied to describe their binding properties. In this picture, the proton and neutron are schematically represented as
\[
p\sim [ud]u,\qquad n\sim [ud]d ,
\]
where the remaining valence \(u\) and \(d\) quarks can be shared and exchanged between neighboring nucleons. The scalar diquark \([ud]_{\bar{\mathbf{3}}_c,I=0,J^P=0^+}\) is antisymmetric in color, flavor, and spin, while its spatial wave function is symmetric. In the present schematic picture, this compact scalar diquark acts as a core constituent analogous to the charm quark in a heavy hadron, and the overlap or exchange of such cores is encoded in the effective repulsive term \(R\).

The resulting effective bond strengths are
\begin{equation}
S\sim 23~{\rm MeV},\qquad
W\sim 17~{\rm MeV},\qquad
\Lambda\sim 18~{\rm MeV},
\end{equation}
while the remaining parameters are estimated as
\begin{equation}
R\sim 14~{\rm MeV},\qquad
\epsilon\sim 4~{\rm MeV}.
\end{equation}
These parameters are taken as transferable phenomenological inputs within the present covalent-bond framework and are applied consistently to hadronic, nuclear, and hypernuclear systems.

Using these inputs, we estimate the generalized binding energies of representative systems within the present framework, with the results summarized in Table~\ref{tab:results}. Additional examples and applications can be found in Refs.~\cite{Chen:2021xlu,Chen:2025jgi}. For each system, the generalized binding energy is obtained by counting the strong, weak, and strange covalent bonds together with the corresponding repulsive exchange contributions according to Eq.~\eqref{eq:Bcov}. For example, the \(T_{cc}(3875)\) state contains one strong covalent bond and one repulsive contribution from heavy-antiquark exchange. Its binding energy is therefore estimated as
\begin{equation}
B_{\rm cov}(T_{cc})
=
1\times S
+
0\times W
+
0\times \Lambda
-
1\times R
-
2\times\epsilon
\sim
1~{\rm MeV}.
\end{equation}
The same bond-counting procedure can be applied to systems composed of a heavy antimeson and a heavy baryon. For the \(P_c(4312)\), treated here as a \(\bar D\Sigma_c\) covalent molecule, the heavy-antiquark exchange contribution present in \(T_{cc}(3875)\) is absent, because the two clusters contain a charm quark and a charm antiquark rather than two identical charm antiquarks. Its binding energy is therefore estimated as
\begin{equation}
B_{\rm cov}(P_c)
=
1\times S
+
0\times W
+
0\times \Lambda
-
0\times R
-
2\times\epsilon
\sim
15~{\rm MeV}.
\end{equation}
The binding energies of light nuclear and hypernuclear systems can be evaluated within the same framework. For the \(^{3}\mathrm{He}\) nucleus, the corresponding covalent-bond counting gives
\begin{equation}
B_{\rm cov}({}^{3}{\rm He})
=
2\times S
+
1\times W
+
0\times \Lambda
-
3\times R
-
3\times\epsilon
\sim
9~{\rm MeV}.
\end{equation}
For the hypernucleus \({}^{3}_{\Lambda}\mathrm{H}\), the strange covalent bond is additionally included, leading to
\begin{equation}
B_{\rm cov}({}^{3}_{\Lambda}{\rm H})
=
1\times S
+
0\times W
+
1\times \Lambda
-
2\times R
-
3\times\epsilon
\sim
1~{\rm MeV}.
\end{equation}
These examples illustrate that the same set of phenomenological bond parameters can be applied consistently across meson--meson, meson--baryon, nuclear, and hypernuclear systems. The resulting generalized binding energies should be regarded as benchmark estimates of the characteristic binding scales rather than precision predictions.

\subsection{Creation Bond}
\label{subsec:creation}

The creation bond was introduced in Ref.~\cite{Chen:2025jgi} to describe near-threshold systems in which the valence light-quark content alone does not permit the direct covalent-exchange pattern discussed in the previous subsection. A representative example is the \(Z_c(3900)\), which is treated in the present framework as a near-threshold \(D\bar D^*\) structure with \(I^G(J^{PC})=1^+(1^{+-})\). For brevity, \(D\bar D^*\) is used throughout to denote the appropriate coherent combination of the \(D\bar D^*\) and \(D^*\bar D\) channels consistent with these quantum numbers. In contrast to the \(\bar D\bar D^*[0(1^+)]\) configuration of \(T_{cc}(3875)\), in which the two mesonic clusters contain two light quarks that can be directly shared and exchanged, the valence light degrees of freedom in the \(D\bar D^*\) system consist of a light quark and a light antiquark. The valence configuration alone therefore does not provide a pair of identical light quarks or antiquarks that can undergo direct covalent exchange, and additional light degrees of freedom must participate in order to generate effective covalent correlations.

One possible realization of this mechanism was proposed in Ref.~\cite{Chen:2025jgi}, where a coherent sea four-quark component, \([q\bar q q\bar q]\), is activated and participates in the interhadronic interaction. This specific arrangement need not be unique, and other realizations involving sea degrees of freedom may also contribute. A more complete understanding of the microscopic origin and dynamics of the creation bond requires further investigation. In the present exploratory study, we adopt the sea four-quark configuration proposed in Ref.~\cite{Chen:2025jgi} as a benchmark realization of the creation-bond mechanism.

As illustrated schematically in Fig.~\ref{fig:bonds}(b), the activated sea four-quark component supplies the additional light-quark degrees of freedom needed to generate effective covalent correlations between the two color-singlet hadronic clusters, which may be viewed collectively as a spring-like link between them. Here, ``activation'' does not imply the literal creation of independent on-shell \(q\bar q\) pairs. Rather, it denotes the participation of the coherent sea four-quark component in the correlated two-hadron configuration. The corresponding net energy cost is denoted by \(E_{\rm act}\) and is assigned to the entire sea four-quark component rather than to each \(q\bar q\) pair separately.

The effective covalent correlations generated through this mechanism are accompanied by an additional color-suppression factor. In the simple color-counting picture adopted here, the relevant light quarks or antiquarks must carry the same color or anticolor. The benchmark sea four-quark configuration involves three such color matchings, giving a combined color-matching probability of \((1/3)^3=1/27\). However, the covalent-bond strengths \(S\), \(W\), and \(\Lambda\) already implicitly incorporate one basic color-matching factor of \(1/3\). To avoid double counting this common factor, only the additional suppression is included. We therefore define
\begin{equation}
\eta_C
=
\frac{(1/3)^3}{1/3}
=
\frac{1}{9},
\label{eq:etaC}
\end{equation}
which represents the additional color suppression relative to the direct covalent bond.

The creation-bond contribution is also sensitive to the spin decomposition of the participating light-quark degrees of freedom. For the \(D\bar D^*\) molecular configurations, the spin-isospin combinations \((I_{q\bar q},S_{q\bar q})=(1,0)\) and \((0,1)\) were identified in Ref.~\cite{Chen:2025jgi} as potentially favorable components, where \(I_{q\bar q}\) and \(S_{q\bar q}\) denote the total isospin and spin of the valence light \(q\bar q\) pair residing in the two mesons. The weights of these components in a given physical channel further depend on the corresponding heavy-quark-spin decomposition~\cite{Isgur:1989vq,Neubert:1993mb,Manohar:2000dt,Guo:2009id,Bondar:2011ev,Hidalgo-Duque:2013pva,Ozpineci:2013zas}. In the present toy model, we do not repeat the detailed channel-by-channel analysis, but directly adopt the favorable bond-counting scheme of Ref.~\cite{Chen:2025jgi}. Moreover, to avoid introducing an additional free parameter, no separate spin-rearrangement contribution is included in the present parametrization. This simplification is motivated by the fact that the relevant spin symmetries are already partially encoded in the distinction between the strong and weak covalent bonds, and by the expectation that the additional color suppression \(\eta_C=1/9\) renders the residual spin-dependent corrections subleading.

Accordingly, the generalized binding energy associated with the creation-bond mechanism is parameterized as
\begin{equation}
B_{\rm cre}
=
\eta_C
\left(
N_S S
+
N_W W
+
N_\Lambda \Lambda
\right)
-
E_{\rm act}
-
N_R R
-
N_h \epsilon ,
\label{eq:Bcre}
\end{equation}
where \(N_S\), \(N_W\), and \(N_\Lambda\) count the effective strong, weak, and strange covalent bonds formed upon activation of the sea four-quark component. The quantities \(N_R\), \(N_h\), \(S\), \(W\), \(\Lambda\), \(R\), and \(\epsilon\) retain the same meanings as in Eq.~\eqref{eq:Bcov}, while \(\eta_C\) and \(E_{\rm act}\) are defined above.

For the \(Z_c(3900)\), we denote the valence light \(q\bar q\) pair by \([q_2\bar q_1]\) and the activated sea four-quark component by \([q_4\bar q_3 q_6\bar q_5]\). Together, they form a configuration containing three light quarks and three light antiquarks. Following the construction proposed in Ref.~\cite{Chen:2025jgi}, the first three rows below specify the color, flavor, spin, and orbital assignments of the three \(q\bar q\) pairs, whereas the remaining six rows list the pairwise covalent correlations among the three light quarks and, separately, among the three light antiquarks:
\begin{equation}\nonumber
\begin{array}{cccccc}
\hline
\textit{creation}
& {\rm color}
& {\rm flavor}
& {\rm spin}
& {\rm orbital}
& {\rm total}
\\ \hline
\bar q_1 \leftrightarrow q_2
& \mathbf{1}_c
& {\bf S}
& {\bf A}
& {\bf S}
& {}^{1}S_0
\\
\bar q_3 \leftrightarrow q_4
& \mathbf{8}_c
& {\bf A}
& {\bf S}
& {\bf S}
& {}^{3}S_1 (\sim gluon)
\\
\bar q_5 \leftrightarrow q_6
& \mathbf{8}_c
& {\bf A}
& {\bf S}
& {\bf S}
& {}^{3}S_1 (\sim gluon)
\\ \hline
\bar q_1 \leftrightarrow \bar q_3
& {\bf S}
& {\bf S}
& {\bf A}
& {\bf S}
& {\bf A}\;(\textit{weak})
\\
\bar q_1 \leftrightarrow \bar q_5
& {\bf S}
& {\bf A}
& {\bf S}
& {\bf S}
& {\bf A}\;(\textit{strong})
\\
\bar q_3 \leftrightarrow \bar q_5
& {\bf S}
& {\bf A}
& {\bf S}
& {\bf S}
& {\bf A}\;(\textit{strong})
\\
q_2 \leftrightarrow q_4
& {\bf S}
& {\bf A}
& {\bf S}
& {\bf S}
& {\bf A}\;(\textit{strong})
\\
q_2 \leftrightarrow q_6
& {\bf S}
& {\bf S}
& {\bf A}
& {\bf S}
& {\bf A}\;(\textit{weak})
\\
q_4 \leftrightarrow q_6
& {\bf S}
& {\bf A}
& {\bf S}
& {\bf S}
& {\bf A}\;(\textit{strong})
\\ \hline
\end{array}
\end{equation}
These six covalent correlations consist of four strong bonds and two weak bonds, yielding
\begin{equation}
N_S=4,\qquad
N_W=2,\qquad
N_\Lambda=0.
\end{equation}
No repulsive heavy-quark exchange contribution arises because the two hadronic clusters contain a charm quark and a charm antiquark rather than two identical heavy constituents. With \(N_R=0\) and \(N_h=2\), Eq.~\eqref{eq:Bcre} gives
\begin{align}
B_{\rm cre}\bigl(Z_c(3900)\bigr)
&=
\frac{1}{9}
\left(
4\times S
+
2\times W
\right)
-
E_{\rm act}
-
2\times\epsilon
\nonumber\\
&\sim
6~{\rm MeV}
-
E_{\rm act}.
\label{eq:BZc_Eact}
\end{align}
Using the experimental generalized binding energy \(B_{\rm exp}\bigl(Z_c(3900)\bigr)\sim-11~{\rm MeV}\), evaluated relative to the isospin-averaged \(D\bar D^*\) threshold, we obtain
\begin{equation}
E_{\rm act}
\sim
17~{\rm MeV}.
\label{eq:Eact}
\end{equation}
The \(Z_c(3900)\) therefore serves as the calibration input for \(E_{\rm act}\), whereas applications to related systems provide phenomenological tests of the creation-bond picture.

Interestingly, the activation energy extracted from the \(Z_c(3900)\) data is numerically close to the reported mass of the \(X_{17}\) particle~\cite{Krasznahorkay:2015iga}:
\begin{equation}
E_{\rm act}
\sim
M_{X17}
=
16.70
\pm 0.35
\pm 0.50
~{\rm MeV}.
\label{eq:EactX17}
\end{equation}
In Ref.~\cite{Chen:2020arr}, the \(X_{17}\) particle was interpreted as a bare four-quark configuration composed of two chiral \(u/d\) quarks and two chiral \(u/d\) antiquarks. Motivated by this picture, we tentatively associate the coherent sea four-quark component involved in the creation bond with a similar low-energy bare four-quark configuration. The activation energy \(E_{\rm act}\) may then be regarded as an effective nonperturbative energy cost for accessing such a configuration. Since the resulting covalent attraction is color suppressed, it need not fully compensate this activation cost, thereby naturally allowing a negative generalized binding energy and an above-threshold structure. Nevertheless, \(E_{\rm act}\) and \(M_{X17}\) have different physical meanings. Their numerical proximity neither implies that an on-shell \(X_{17}\) state is produced nor establishes a microscopic identification between them. Rather, it may hint at a common low-energy scale associated with confinement-related nonperturbative dynamics. The scale of \(17~{\rm MeV}\), however, should not be identified with the fundamental confinement scale itself.

The same creation-bond mechanism may also be relevant to other hidden-heavy and open-flavor systems, including the \(Z_c(4020)\), \(Z_b(10610)\), \(Z_b(10650)\), \(T_{c \bar s 0}^a(2900)\), \(Z_{cs}(3985)\), and \(Z_{cs}(4000)\) candidates listed in Table~\ref{tab:results}. Their detailed threshold positions may depend on the heavy-flavor content, spin configuration, strange-quark contributions, and the microscopic realization of the activated sea four-quark component. The present framework is intended to characterize the common creation-bond features and associated generalized-binding-energy scales of these systems, rather than to provide precision descriptions of their individual masses or line shapes.

\subsection{Annihilation Bond}
\label{subsec:annihilation}

The \(X(3872)\) is particularly distinctive among the near-threshold systems considered in the present framework. Under the benchmark bond counting adopted above, the creation-bond contribution alone would place the corresponding \(D\bar D^*\) configuration above the isospin-averaged threshold, whereas the observed \(X(3872)\) lies slightly below it. This suggests the presence of an additional mechanism. Besides the activation of the coherent sea four-quark component considered in the creation bond, the valence light quark and valence light antiquark residing in the two hadronic clusters can also annihilate. Since quark-pair creation and annihilation are complementary quark-line processes, we refer to the effective contribution induced by valence \(q\bar q\) annihilation as the annihilation bond.

The annihilation bond is not regarded as an independent binding mechanism within the present framework, but rather as a modification of a pre-existing two-hadron configuration already correlated through either the covalent or creation bond. As illustrated schematically in Fig.~\ref{fig:bonds}(c), valence \(q\bar q\) annihilation changes the quark-line topology and tends to destabilize the original two-hadron configuration by depleting its molecular component. In the absence of a nearby compact state with the same quantum numbers, the annihilation process may predominantly broaden or dissolve the molecular component, thereby preventing it from developing into an experimentally identifiable near-threshold structure~\cite{Chen:2023zlj}.

A coherent annihilation contribution can arise when a nearby compact state with the same quantum numbers is available, allowing the molecular configuration to mix with the compact state. Such mixing produces two shifted physical eigenstates, but they need not both be experimentally identifiable. Depending on the mixing pattern and the relevant production and decay couplings, the experimentally observed state may be predominantly molecular or predominantly compact. Thus, the molecular correlation need not increase the number of experimentally identifiable states, but may instead reorganize the nearby spectrum.

Schematically, in the basis spanned by the unperturbed two-hadron configuration and a nearby bare compact state, the annihilation-induced mixing may be represented by the mass matrix
\begin{equation}
\mathcal{M}_{\rm ann}=
\begin{pmatrix}
M_{\rm cov/cre} & V_{\rm ann}\\
V_{\rm ann} & M_{\rm compact}
\end{pmatrix},
\label{eq:Mann}
\end{equation}
where \(M_{\rm compact}\) denotes the mass of this compact state and \(V_{\rm ann}\) represents the effective mixing induced by valence \(q\bar q\) annihilation. The quantity \(M_{\rm cov/cre}\) denotes the mass of the unperturbed two-hadron configuration,
\begin{equation}
M_{\rm cov/cre}=M_{\rm threshold}-B_{\rm cov/cre},
\label{eq:Mcovcre}
\end{equation}
where \(B_{\rm cov/cre}\) denotes its generalized binding energy before annihilation-induced mixing. Here and below, the notation ``cov/cre'' refers to either a covalent-bond configuration or a creation-bond configuration, as appropriate.

A detailed treatment would require information on the relevant compact state and the corresponding mixing matrix element. In the present toy model, we focus on the eigenstate continuously connected to the unperturbed two-hadron configuration and encode the net mixing effect in a single signed parameter,
\begin{equation}
\Delta=M_{\rm cov/cre}-M,
\label{eq:Delta}
\end{equation}
where \(M\) denotes the mass of this physical eigenstate. Its generalized binding energy after including annihilation-induced mixing is written as
\begin{equation}
B_{\rm ann}=M_{\rm threshold}-M=B_{\rm cov/cre}+\Delta.
\label{eq:Bann}
\end{equation}
A positive value of \(\Delta\) lowers the physical mass and increases the generalized binding energy, whereas a negative value raises the mass and decreases the generalized binding energy. In the weak-mixing limit and away from exact degeneracy,
\begin{equation}
\Delta\simeq\frac{V_{\rm ann}^{\,2}}{M_{\rm compact}-M_{\rm cov/cre}},
\label{eq:DeltaMixing}
\end{equation}
showing that the sign of the mass shift depends on the relative positions of the unperturbed two-hadron configuration and the compact state. Thus, the annihilation bond can shift the physical eigenstate either downward or upward in mass and need not always enhance its binding.

For the \(X(3872)\), the relevant compact state is taken to be the nearby bare \(\chi_{c1}(2P)\) charmonium state. This state is assumed to lie above the unperturbed \(D\bar D^*\) configuration, so that level repulsion shifts the predominantly molecular eigenstate downward. The \(X(3872)\) has the same benchmark creation-bond counting as the \(Z_c(3900)\). After including the annihilation-induced contribution, we obtain from Eq.~\eqref{eq:BZc_Eact}:
\begin{align}
B_{\rm ann}\bigl(X(3872)\bigr)
&=
\frac{1}{9}
\left(
4\times S
+
2\times W
\right)
-
E_{\rm act}
-
2\times\epsilon
+
\Delta
\nonumber\\
&\sim
-11~{\rm MeV}
+
\Delta.
\label{eq:BX3872}
\end{align}
Using the experimental generalized binding energy \(B_{\rm exp}\bigl(X(3872)\bigr)\sim4~{\rm MeV}\), evaluated relative to the isospin-averaged \(D\bar D^*\) threshold, we obtain
\begin{equation}
\Delta\sim15~{\rm MeV}.
\label{eq:DeltaX}
\end{equation}
The \(X(3872)\) therefore serves as the calibration input for the characteristic magnitude of the annihilation-induced contribution. In the benchmark estimates presented in Table~\ref{tab:results}, we adopt
\begin{equation}
|\Delta|\sim15~{\rm MeV},
\end{equation}
while allowing its sign to depend on the relative positions of the unperturbed molecular configuration and the compact state.

As an application, the \(X(3960)\)~\cite{LHCb:2022aki} is treated as a \(D_s\bar D_s\) configuration containing both creation- and annihilation-bond contributions. The relevant compact state is taken to be the nearby bare \(\chi_{c0}(2P)\) charmonium state, assumed to lie below the unperturbed \(D_s\bar D_s\) configuration. Level repulsion then shifts the predominantly molecular eigenstate upward, corresponding to the benchmark value \(\Delta=-15~{\rm MeV}\). Using \(N_S=2\), \(N_\Lambda=2\), \(N_h=2\), and \(N_W=N_R=0\), we obtain
\begin{align}
B_{\rm ann}\bigl(X(3960)\bigr)
&=
\frac{1}{9}
\left(
2\times S
+
2\times\Lambda
\right)
-
E_{\rm act}
-
2\times\epsilon
+
\Delta
\nonumber\\
&\sim
-31~{\rm MeV}.
\label{eq:BX3960}
\end{align}
The resulting negative generalized binding energy places the predominantly molecular eigenstate above the \(D_s\bar D_s\) threshold and is consistent with the experimental value in sign and order of magnitude. It should nevertheless be regarded as a benchmark estimate rather than a precision description.

The annihilation bond therefore parametrizes the mass shift induced when valence \(q\bar q\) annihilation couples an already correlated two-hadron configuration to a nearby compact state. Within the present framework, it is not an independent binding mechanism, but acts together with either the covalent or creation bond as a signed correction that can enhance or reduce the generalized binding energy.

\section{Analysis and Applications}
\label{sec:analysis}

The hadronic-bond framework introduced in the previous section is deliberately minimal. At the benchmark level, the same set of phenomenological bond parameters is applied across different flavor sectors and hadronic configurations, allowing the near-threshold systems summarized in Table~\ref{tab:results} to be organized according to common light-quark processes. In this section, we extract the general binding patterns and model predictions that emerge from this unified classification, and explore several extensions of the basic bond-counting picture.

We first summarize the main patterns exhibited by the covalent, creation, and annihilation bonds. We then compare the covalent and creation mechanisms, followed by a discussion of the complementary roles of creation and annihilation. Finally, we extend the bond-counting picture to multihadron systems and explore the more complicated behavior of kaon-related near-threshold systems.

\subsection{General Patterns}
\label{subsec:general_patterns}

Several simple patterns emerge from the systems collected in Table~\ref{tab:results}. First, the configurations supported primarily by covalent bonds are generally located below their corresponding two-hadron thresholds, whereas those supported primarily by creation bonds are generally located above thresholds. In terms of the generalized binding energy, this tendency may be summarized schematically as
\begin{equation}
B_{\rm cov}>0,\qquad B_{\rm cre}<0,
\label{eq:threshold_pattern}
\end{equation}
for most of the representative systems considered here. The covalent bond directly utilizes the valence light-quark degrees of freedom already present in the two hadronic clusters and therefore tends to produce below-threshold bound states. By contrast, the attractive correlation supplied by the creation bond is reduced by both the color-suppression factor \(\eta_C\) and the activation energy \(E_{\rm act}\), so that the resulting structures predominantly remain above threshold and appear experimentally as resonance candidates.

The annihilation bond is more restrictive because it acts only on a pre-existing correlated two-hadron configuration containing an available valence \(q\bar q\) pair. In addition, an experimentally identifiable state shifted by this mechanism may require mixing with a nearby compact state. These conditions naturally reduce the number of candidate systems in which annihilation plays an essential role. Its characteristic effect is therefore not to generate a large family of states, but to induce a signed mass shift of an existing correlated configuration. As expressed in Eq.~\eqref{eq:Bann}, a positive value of \(\Delta\) increases the generalized binding energy, whereas a negative value decreases it.

Beyond these bond-mechanism patterns, core repulsion provides another source of systematic variation. Experimentally, the \(P_c(4312)\) has a generalized binding energy of approximately \(9~{\rm MeV}\), larger than those of the \(T_{cc}(3875)\) and the deuteron, which are approximately \(1\) and \(2~{\rm MeV}\), respectively. In the present core picture, the \(T_{cc}\) system contains two identical heavy-quark cores, while the proton and neutron in the deuteron each contain a compact \([ud]\) core. By contrast, the \(\bar D\Sigma_c\) configuration assigned to the \(P_c(4312)\) contains distinguishable \(\bar c\) and \(c\) cores, so that no exchange-related core-repulsion term is included. The comparatively stronger binding of the \(P_c(4312)\) therefore provides a qualitative indication that the absence of identical-core repulsion can significantly enhance the binding, although differences in the light-quark configurations and other channel-dependent effects may also contribute. For otherwise analogous covalent-bond configurations and before accounting for the heavy-core-mass dependence of \(\epsilon\), applying the same reasoning to the bottom--charm system gives
\begin{equation}
B_{\rm cov}(T_{bc})>B_{\rm cov}(T_{cc})\sim B_{\rm cov}(T_{bb}),
\label{eq:Tbc_core}
\end{equation}
because \(T_{bc}\) contains distinguishable \(b\) and \(c\) cores, whereas \(T_{cc}\) and \(T_{bb}\) each contain two identical heavy cores.

The comparison between the charm and bottom creation-bond systems provides a possible indication of the heavy-core-mass dependence of the kinetic correction. As shown in Table~\ref{tab:results}, the generalized binding energies of the \(Z_b(10610)\) and \(Z_b(10650)\) structures are both approximately \(-3~{\rm MeV}\), compared with approximately \(-11\) and \(-7~{\rm MeV}\) for the \(Z_c(3900)\) and \(Z_c(4020)\), respectively. The less negative values in the bottom sector suggest a stronger tendency toward binding and may be phenomenologically encoded as
\begin{equation}
\epsilon_b<\epsilon_c.
\label{eq:epsilon_heavy}
\end{equation}
If the same heavy-core-mass dependence applies to otherwise analogous covalent-bond configurations, \(T_{bb}\) is expected to be more tightly bound than \(T_{cc}\). Combining this effect with the absence of identical-core repulsion in \(T_{bc}\) leads to the qualitative ordering
\begin{equation}
B_{\rm cov}(T_{bc})>B_{\rm cov}(T_{bb})>B_{\rm cov}(T_{cc}).
\label{eq:Tbc_prediction}
\end{equation}
This ordering thus constitutes a distinctive and directly testable prediction of the present framework.

\subsection{Covalent versus Creation Bonds}
\label{subsec:cov_creation}

The covalent and creation bonds should not be regarded as two unrelated sources of attraction. Both mechanisms ultimately generate effective correlations through the sharing and exchange of light-quark degrees of freedom between two color-singlet hadronic clusters. Their essential difference lies in the origin and connectivity of the active light constituents. In a covalent-bond configuration, the relevant light quarks are already present in the valence components of the two hadrons, so that the resulting correlation resembles a direct clasp between the two clusters. In a creation-bond configuration, the required light degrees of freedom are instead supplied by an activated sea component, producing a multilayer spring-like linkage formed through several intermediate light-quark connections. The creation bond may therefore be viewed as a sea-assisted extension of the covalent-bond mechanism.

The distinction is illustrated by the comparison between the \(T_{cc}(3875)\) and \(Z_c(3900)\) systems, interpreted in the present study as \(DD^*\) and \(D\bar D^*\) configurations, respectively. In the \(DD^*\) configuration, the valence light quarks supplied by the two hadronic clusters can directly form one strong covalent bond. By contrast, the \(D\bar D^*\) configuration requires the activation of additional sea-quark degrees of freedom before analogous correlations can be established. Although the activated component produces four strong and two weak bond contributions, their net effect is suppressed by the color factor \(\eta_C=1/9\) and reduced by the activation energy \(E_{\rm act}\). Consequently, the direct covalent bond tends to support a below-threshold bound state, whereas the more indirect and energetically costly creation bond generally favors an above-threshold resonance candidate. The contrast between the two mechanisms therefore lies not in attraction versus repulsion, but in direct valence bonding versus sea-assisted bonding.

A further structural relation is suggested by the \(T_{\bar c\bar s0}^{*}(2870)\)~\cite{LHCb:2020bls,LHCb:2024xyx} and \(T_{c\bar s0}^{a}(2900)\)~\cite{LHCb:2022sfr} systems, interpreted in the present study as \(\bar D^{*}K^{*}\) and \(D^{*}K^{*}\) configurations, respectively. When the \(c/\bar c\) and \(\bar s\) constituents are treated schematically as cores, the remaining active light degrees of freedom form a \(qq\) pair in the \(\bar D^{*}K^{*}\) configuration and a \(q\bar q\) pair in the \(D^{*}K^{*}\) configuration. The former can therefore support a direct covalent bond, whereas the latter requires a sea-assisted creation bond.

A particularly clean prediction follows from comparing the \(P_{cs}(4459)\) with its hypothetical isovector \(D^*\Xi_c\) counterpart. The \(P_{cs}(4459)\) is interpreted as an isoscalar \(\bar D^*\Xi_c\) configuration with an experimental generalized binding energy of approximately \(19~{\rm MeV}\). In the \(\bar D^*\Xi_c\sim(\bar c q)(csq)\) configuration, the active \(qq\) pair forms one strong covalent bond, while the distinguishable \(\bar c\) and \(c\) cores generate no identical-core repulsion. Eq.~\eqref{eq:Bcov} then gives
\begin{equation}
B_{\rm cov}\bigl(P_{cs}(4459)\bigr)
=
1\times S
-
2\times\epsilon
\sim
15~{\rm MeV},
\label{eq:BPcs}
\end{equation}
in reasonable agreement with the experimental value. By contrast, the \(D^*\Xi_c\sim(c\bar q)(csq)\) configuration is supported by a creation bond and receives one identical-core repulsive contribution. For \(I=1\), annihilation-induced mixing with a compact \(ccs\) state is forbidden by isospin conservation, so that \(\Delta=0\). Eq.~\eqref{eq:Bcre} therefore gives
\begin{align}
&B_{\rm cre}\bigl(D^*\Xi_c[I=1]\bigr)
\nonumber\\
&~~~~~~~~=
\frac{1}{9}
\left(
4\times S+2\times W
\right)
-
E_{\rm act}
-
1\times R
-
2\times\epsilon
\nonumber\\
&~~~~~~~~\sim
-25~{\rm MeV}.
\label{eq:BDXi}
\end{align}
The isovector \(D^*\Xi_c\) system is thus predicted to lie substantially above threshold, in sharp contrast to the bound \(P_{cs}(4459)\), providing a clean test of the covalent--creation distinction.

\subsection{Creation--Annihilation Complementarity}
\label{subsec:creation_annihilation}

The introduction of the creation bond naturally calls for a complementary annihilation mechanism: if the nonperturbative QCD vacuum can activate an additional \(q\bar q\) pair, an available \(q\bar q\) pair in a correlated hadronic configuration can also annihilate and rearrange the system. Such annihilation-induced rearrangement has long been studied in coupled-channel descriptions of the \(X(3872)\), where the physical state is interpreted as a mixture of a compact \(c\bar c\) configuration and the \(D\bar D^*\) component with \(I^G(J^{PC})=0^+(1^{++})\)~\cite{Meng:2005er,Meng:2014ota,Matheus:2009vq,Brambilla:2024imu,Micu:1968mk,LeYaouanc:1972vsx,Barnes:2005pb}. In the present phenomenological picture, pair creation and pair annihilation are regarded as complementary manifestations of vacuum-induced quark rearrangement. Both may be associated with color confinement and have no direct analogue in ordinary electromagnetic molecular bonding.

Their roles in the hadron spectrum are nevertheless different. A creation bond establishes the interhadronic correlation when the required light constituents are absent from the valence configurations, whereas an annihilation bond acts on an already correlated configuration containing an available valence \(q\bar q\) pair and couples it to a nearby compact state. Such mixing requires compatible quantum numbers, and its spectral effect depends on the relative positions of the two configurations. As expressed in Eqs.~\eqref{eq:Bann} and \eqref{eq:DeltaMixing}, the resulting contribution \(\Delta\) is signed: a compact state above the unperturbed two-hadron configuration shifts the corresponding eigenstate downward, whereas one below it shifts the eigenstate upward. Annihilation can therefore either enhance or reduce the generalized binding energy.

The \(Z_c(3900)\) and \(X(3872)\), together with a possible \(X(3960)\) assignment, illustrate this complementarity. The \(Z_c(3900)\) and \(X(3872)\) are both associated with \(D\bar D^*\) configurations and share the same benchmark creation-bond counting. The \(Z_c(3900)\), which remains above threshold, calibrates the activation energy \(E_{\rm act}\) and defines the creation-bond baseline. For the \(X(3872)\), the nearby \(\chi_{c1}(2P)\) compact state is assumed to lie above the unperturbed \(D\bar D^*\) configuration, so that annihilation-induced mixing produces a positive contribution, \(\Delta\sim+15~{\rm MeV}\), and shifts the predominantly \(D\bar D^*\) eigenstate below threshold. The \(X(3960)\) may illustrate the opposite behavior if it is assigned to a \(D_s\bar D_s\) configuration coupled to the nearby \(\chi_{c0}(2P)\) compact state. With the compact state assumed to lie below the unperturbed \(D_s\bar D_s\) configuration, level repulsion shifts the eigenstate connected to the two-hadron configuration upward, yielding \(\Delta\sim-15~{\rm MeV}\).

The same creation--annihilation complementarity may extend naturally to baryon--antibaryon systems. A close analogue of the \(Z_c(3900)\)--\(X(3872)\) comparison is provided by the \(\Xi_c\bar\Xi_c\sim(csq)(\bar c\bar s\bar q)\) configurations with \(I=1\) and \(I=0\). For channels with the same \(J^{PC}\), the two configurations share the same hadronic constituents, threshold, and benchmark creation-bond counting. Their total isospin is carried by the nonstrange \(q\bar q\) pair, whose annihilation is forbidden for \(I=1\) but allowed for \(I=0\) in the isospin-symmetric limit. The isoscalar channel may therefore receive an additional signed mass shift, with its benchmark magnitude estimated as
\begin{align}
&\left|
B_{\rm cre}\bigl(\Xi_c\bar\Xi_c[I=1]\bigr)
-
B_{\rm ann}\bigl(\Xi_c\bar\Xi_c[I=0]\bigr)
\right|
\nonumber\\
&~~~~~~~~~~~~~~~~~~~~~~~~~~~~~~~~~~~~~~~~~\simeq
|\Delta|
\sim
15~{\rm MeV},
\label{eq:XiXi_ann}
\end{align}
provided that a nearby compact state with compatible \(J^{PC}\) exists and any common \(s\bar s\)-annihilation contribution is neglected. The sign of the mass shift depends on whether this compact state lies above or below the unperturbed \(\Xi_c\bar\Xi_c\) configuration. A splitting of this characteristic magnitude would provide a distinctive and directly testable signature of the present framework.

\subsection{Multihadron Systems}
\label{subsec:multihadron}

Light nuclei provide the most direct guide for extending the covalent-bond picture from two-hadron configurations to multihadron systems. In the schematic core representation \(p\sim[ud]u\) and \(n\sim[ud]d\), the compact scalar diquark \([ud]\) acts as the core, while the remaining valence light quarks can be shared and exchanged among neighboring nucleons. The essential point is that an active light quark is not restricted to a single localized bond. This is illustrated by the schematic pairwise symmetry assignments introduced in Ref.~\cite{Chen:2021xlu}:
\begin{itemize}

\item At the level of the shared light degrees of freedom, the deuteron realizes the same one-strong-bond pattern as the \(T_{cc}(3875)\). Its two shared light quarks form one strong bond:
\begin{equation}\nonumber
\begin{array}{cccccc}
\hline
{}^{2}{\rm H}                & {\rm color} & {\rm flavor} & {\rm spin} & {\rm orbital} & {\rm total}
\\ \hline
q_1 \leftrightarrow q_2      & {\bf S}     & {\bf A}      & {\bf S}    & {\bf S}       & {\bf A}\;(\textit{strong})
\\ \hline
\end{array}
\end{equation}

\item In the \(A=3\) nuclei, the three shared light quarks form a connected three-center configuration containing two strong bonds and one weak bond:
\begin{equation}\nonumber
\begin{array}{cccccc}
\hline
{}^{3}{\rm H}/{}^{3}{\rm He} & {\rm color} & {\rm flavor} & {\rm spin} & {\rm orbital} & {\rm total}
\\ \hline
q_1 \leftrightarrow q_2      & {\bf S}     & {\bf A}      & {\bf S}    & {\bf S}       & {\bf A}\;(\textit{strong})
\\
q_1 \leftrightarrow q_3      & {\bf S}     & {\bf A}      & {\bf S}    & {\bf S}       & {\bf A}\;(\textit{strong})
\\
q_2 \leftrightarrow q_3      & {\bf S}     & {\bf S}      & {\bf A}    & {\bf S}       & {\bf A}\;(\textit{weak})
\\ \hline
\end{array}
\end{equation}
Within the present schematic picture, the resulting structure is therefore not a deuteron-like pair accompanied by a spectator nucleon.

\item The multicenter character is even more pronounced in \(^{4}{\rm He}\), where the four shared light quarks form a highly connected four-center network containing four strong and two weak bonds:
\begin{equation}\nonumber
\begin{array}{cccccc}
\hline
{}^{4}{\rm He}               & {\rm color} & {\rm flavor} & {\rm spin} & {\rm orbital} & {\rm total}
\\ \hline
q_1 \leftrightarrow q_2      & {\bf S}     & {\bf S}      & {\bf A}    & {\bf S}       & {\bf A}\;(\textit{weak})
\\
q_1 \leftrightarrow q_3      & {\bf S}     & {\bf A}      & {\bf S}    & {\bf S}       & {\bf A}\;(\textit{strong})
\\
q_1 \leftrightarrow q_4      & {\bf S}     & {\bf A}      & {\bf S}    & {\bf S}       & {\bf A}\;(\textit{strong})
\\
q_2 \leftrightarrow q_3      & {\bf S}     & {\bf A}      & {\bf S}    & {\bf S}       & {\bf A}\;(\textit{strong})
\\
q_2 \leftrightarrow q_4      & {\bf S}     & {\bf A}      & {\bf S}    & {\bf S}       & {\bf A}\;(\textit{strong})
\\
q_3 \leftrightarrow q_4      & {\bf S}     & {\bf S}      & {\bf A}    & {\bf S}       & {\bf A}\;(\textit{weak})
\\ \hline
\end{array}
\end{equation}

\end{itemize}
Unlike electrons, for which a single spatial orbital can accommodate at most two electrons with opposite spins, light quarks carry color and flavor degrees of freedom in addition to spin. More than two light quarks can therefore participate in a common spatially symmetric multicenter wave function, while the required antisymmetry of the total wave function is ensured by their internal color--flavor--spin structure. Accordingly, the pairwise symmetry assignments listed above should be understood as different correlations within this common antisymmetrized wave function, rather than as a collection of independently optimized two-hadron bonds.

As shown in Table~\ref{tab:results}, the binding energies of \(^{2}{\rm H}\), \(^{3}{\rm H}/{}^{3}{\rm He}\), and \(^{4}{\rm He}\) were estimated to be approximately \(1\), \(9\), and \(26~{\rm MeV}\), respectively, in reasonable agreement with experiment. The sharp increase in the average binding energy per nucleon from approximately \(1\)--\(3~{\rm MeV}\) for the \(A=2\) and \(A=3\) nuclei to approximately \(7~{\rm MeV}\) for \(^{4}{\rm He}\) provides important phenomenological motivation for a collective multicenter bonding picture. Within the present framework, the fourth nucleon opens two additional strong bonds and one additional weak bond, transforming the relatively open three-center configuration into a highly connected four-center network. This cooperative enhancement provides a qualitative explanation for the unusually strong binding of the \(\alpha\) particle, although the observed binding pattern does not by itself uniquely establish light-quark delocalization as its microscopic origin.

The same multicenter mechanism may also operate in exotic systems composed of multiple \(\bar D^*\) mesons. For compatible \(S\)-wave spin couplings in which the shared light-quark subsystem realizes the same spin--flavor symmetry patterns as in the corresponding light nuclei, the benchmark counting gives:
\begin{itemize}

\item The \(\bar D^*\bar D^*\sim(\bar cq)(\bar cq)\) configuration realizes the same one-strong-bond pattern as the deuteron and the \(T_{cc}(3875)\). Its two shared light quarks form one strong bond, leading to a benchmark binding energy of approximately \(1~{\rm MeV}\).

\item The \(\bar D^*\bar D^*\bar D^*\sim(\bar cq)(\bar cq)(\bar cq)\) configuration is analogous to the \(A=3\) nuclei. Its three shared light quarks form a connected three-center configuration containing two strong bonds and one weak bond, leading to a benchmark binding energy of approximately \(9~{\rm MeV}\).

\item The \(\bar D^*\bar D^*\bar D^*\bar D^*\sim(\bar cq)(\bar cq)(\bar cq)(\bar cq)\) configuration is analogous to the \(^{4}{\rm He}\) nucleus. Its four shared light quarks form a highly connected four-center network containing four strong and two weak bonds, leading to a benchmark binding energy of approximately \(26~{\rm MeV}\).

\end{itemize}

A second example is provided by the isoscalar \(\bar D^*\bar D^*\Sigma_c\sim(\bar c q_1)(\bar c q_2)(c q_3q_4)\) configuration. The pairs \(q_1q_3\) and \(q_2q_4\) can form two strong bonds, while \(q_1q_2\) forms one weak bond in the channel considered. At first sight, the corresponding color requirements appear incompatible: the exchange contribution associated with \(q_1q_3\) requires a same-color component for the first strong bond, and the same holds for \(q_2q_4\) in the second, whereas \(q_3q_4\) are antisymmetric in color inside the \(\Sigma_c\). In a fixed color component satisfying both strong-exchange conditions, \(q_1\) and \(q_2\) would then carry different colors and could not form the weak bond. However, the individual bond contributions are evaluated as pairwise exchange matrix elements in the normalized total color-singlet wave function, rather than by imposing all three exchange conditions simultaneously on a fixed color assignment. The \(q_1q_2\) exchange contribution therefore need not vanish and is retained as one weak bond in the present benchmark counting, although its precise weight would require an explicit spin--flavor and color recoupling calculation. Altogether, we estimate~\footnote{An alternative possibility is that the \(q_1q_2\) exchange is strongly suppressed or effectively absent once full spin--flavor and color recoupling effects are taken into account. The system would then retain only the two strong \(\bar D^*\)--\(\Sigma_c\) links, forming an \(\mathrm{H}_2\mathrm{O}\)-like open three-hadron bond topology rather than a closed triangular network. Within the present picture, if the \(q_1q_2\) exchange is absent, the corresponding exchange-induced repulsion between the two identical \(\bar c\) cores is also expected to be strongly suppressed. Neglecting this repulsive contribution in the benchmark counting, the binding-energy estimate becomes \(B_{\rm cov}=2S-3\epsilon\sim34~{\rm MeV}\). Here the \(\mathrm{H}_2\mathrm{O}\) analogy refers only to the bond connectivity, not to the spatial geometry.}:
\begin{align}
&B_{\rm cov}\bigl(\bar D^*\bar D^*\Sigma_c[I=0]\bigr)
\nonumber\\
&\hspace{1.0cm}=
2\times S
+
1\times W
+
0\times\Lambda
-
1\times R
-
3\times\epsilon
\nonumber\\
&\hspace{1.0cm}\sim
37~{\rm MeV}.
\label{eq:DbarDbarSigmac}
\end{align}
The comparatively large benchmark binding arises from the presence of two strong and one weak attractive bonds, which are opposed by only one exchange-related core-repulsion contribution associated with the two identical \(\bar c\) cores in the present counting scheme.

We further consider the isoscalar \(D^*\bar D^*B^*\bar B^*\) system as an example in which creation and covalent bonds coexist. The experimentally observed \(Z_c(4020)\sim D^*\bar D^*\) and \(Z_b(10650)\sim B^*\bar B^*\) are treated as isovector creation-bond subsystems and subsequently coupled to total isospin \(I=0\). In addition, the \(D^*\bar B^*\) and \(\bar D^*B^*\) pairs may each support a strong covalent bond, leading to a square-like four-center bond topology, while the diagonal \(D^*B^*\) and \(\bar D^*\bar B^*\) interactions are neglected at the present benchmark level. We therefore estimate
\begin{align}
&B\bigl(D^*\bar D^*B^*\bar B^*[I=0]\bigr)
\nonumber\\
&\hspace{1.0cm}\sim
2\times S-7~{\rm MeV}-3~{\rm MeV}
\sim36~{\rm MeV}.
\end{align}
Although this represents only one possible bonding configuration and annihilation effects may further modify the spectrum, the sizable benchmark binding suggests that this system may be rather deeply bound. The \(I=1/2\) \(D\bar D^*B^*\sim[X(3872)+B^*]\) system may likewise support a moderately bound three-hadron configuration.

The multihadron examples discussed above should be regarded as schematic applications of the present bond-counting approximation. For an \(N\)-hadron configuration, the generalized binding energy is defined relative to the corresponding \(N\)-hadron threshold and does not by itself imply stability against lower hadronic rearrangement thresholds. Nevertheless, within the present picture, the light-nuclear examples suggest that more than two light quarks can occupy the same spatial orbital, with the required antisymmetry accommodated by their internal color--flavor--spin degrees of freedom. This allows the participating light constituents to form a connected multicenter network and thereby cooperatively enhance multihadron binding.

\subsection{Kaon-Related Systems}
\label{subsec:kaon_related}

Several kaon-related near-threshold systems, as listed in Table~\ref{tab:krelated}, cannot be straightforwardly accommodated within the present toy model. In this subsection, we examine whether their common features can nevertheless be understood qualitatively within the hadronic-bond framework. We explore the possibility that, owing to its special pseudoscalar nature, the kaon as a whole may participate in the interhadronic covalent correlation. We encode the corresponding additional attraction phenomenologically through an effective covalent-bond contribution, which we continue to denote by \(\Lambda\), and adopt the same benchmark value \(\Lambda\sim18~{\rm MeV}\) introduced in Sec.~\ref{subsec:covalent}. This value is retained only as a common reference scale and need not exhaust the kaon-specific attraction.

We emphasize that this assignment is exploratory because its microscopic origin remains uncertain and requires further investigation. It is motivated by the comparatively low mass and Goldstone nature of the pseudoscalar kaon, together with the near-threshold \(S\)-wave dynamics of the systems considered here. These features may favor a sizable kaon-cloud component and thereby weaken a fixed core--active assignment of the individual kaon constituents. This treatment still permits a valence light \(q\bar q\) pair residing in different hadronic clusters to undergo annihilation when the quantum numbers allow, so that the annihilation-induced correction \(|\Delta|\sim15~{\rm MeV}\) can be incorporated independently of \(\Lambda\). No analogous whole-meson participation is assumed for the heavier vector \(K^*\) in the \(D^*K^*\) systems of Table~\ref{tab:results}, which are treated according to the original core--active assignment.

\begin{table}[t]
\centering
\renewcommand{\arraystretch}{1.5}
\caption{Generalized binding energies of representative kaon-related systems considered in an exploratory extension of the hadronic-bond framework. The pseudoscalar kaon is treated as a whole in forming the interhadronic covalent correlation, and the corresponding effective contribution is parametrized by the benchmark scale \(\Lambda\sim18~{\rm MeV}\). For the isoscalar configurations, a common benchmark annihilation-induced contribution, \(\Delta\sim+15~{\rm MeV}\), is provisionally adopted. The \(T_{c\bar s1}^{a}(2470)\) marked by \(\dagger\) denotes a predicted isovector \(D^*K\) partner of the \(T_{c\bar s0}^{a}(2327)\), with a generalized binding energy of approximately \(36~{\rm MeV}\) inferred through heavy-quark spin symmetry. The recently observed \(B_{s0}^*(5700)^0\)~\cite{LHCb:2026knl} is included under its favored \(J^P=0^+\) assignment as a possible bottom counterpart of the \(D_{s0}^*(2317)\).}
\label{tab:krelated}
\begin{tabular}{c c c c c}
\hline
System
& Configuration
& \(B_{\rm exp}\)
& Counting
& \(B_{\rm calc}\)
\\ \hline
\(T_{c\bar s0}^{a}(2327)\)
& \(DK[1(0^+)]\)
& \(36~{\rm MeV}\)
& \(\Lambda-2\epsilon\)
& \(10~{\rm MeV}\)
\\
\(T_{c\bar s1}^{a}(2470)^{\dagger}\)
& \(D^*K[1(1^+)]\)
& --
& \(\Lambda-2\epsilon\)
& \(10~{\rm MeV}\)
\\
\(D_{s0}^*(2317)\)
& \(DK[0(0^+)]\)
& \(45~{\rm MeV}\)
& \(\Lambda-2\epsilon+\Delta\)
& \(25~{\rm MeV}\)
\\
\(D_{s1}(2460)\)
& \(D^*K[0(1^+)]\)
& \(45~{\rm MeV}\)
& \(\Lambda-2\epsilon+\Delta\)
& \(25~{\rm MeV}\)
\\ 
\(B_{s0}^*(5700)\) 
& \(BK[0(0^+)]\) 
& \(76~{\rm MeV}\) 
& \(\Lambda-2\epsilon+\Delta\) 
& \(25~{\rm MeV}\)
\\
\(\Lambda(1405)\)
& \(\bar KN[0(1/2^-)]\)
& \(30~{\rm MeV}\)
& \(\Lambda-2\epsilon+\Delta\)
& \(25~{\rm MeV}\)
\\
\(\Omega(2012)\)
& \(\bar K\Xi^*[0(3/2^-)]\)
& \(17~{\rm MeV}\)
& \(\Lambda-2\epsilon+\Delta\)
& \(25~{\rm MeV}\)
\\ \hline
\end{tabular}
\end{table}

The comparison between the \(T_{c\bar s0}^{a}(2327)\)~\cite{LHCb:2024iuo} and the \(D_{s0}^*(2317)\) is particularly instructive because the two systems are interpreted as isovector and isoscalar \(DK\) configurations, respectively, and therefore share the same benchmark kaon-assisted covalent contribution. In the isovector channel, valence \(q\bar q\) annihilation cannot couple the \(DK\) configuration to a compact \(c\bar s\) state in the isospin-symmetric limit, leading to the benchmark estimate \(B_{\rm cov}=\Lambda-2\epsilon\sim10~{\rm MeV}\). In the isoscalar channel, such annihilation is allowed and may mix the \(DK\) configuration with a nearby compact \(c\bar s\) state, giving \(B_{\rm ann}=\Lambda-2\epsilon+\Delta\sim25~{\rm MeV}\).

Using the experimental central values, the generalized binding energies of the two systems are approximately \(36\) and \(45~{\rm MeV}\), respectively, suggesting an isoscalar enhancement of approximately
\begin{equation}
B_{\rm exp}\bigl(D_{s0}^*(2317)\bigr)-B_{\rm exp}\bigl(T_{c\bar s0}^{a}(2327)\bigr)\sim9~{\rm MeV}.
\label{eq:DK_isospin}
\end{equation}
Although this difference is not yet significant in view of the experimental uncertainty of the \(T_{c\bar s0}^{a}(2327)\), its sign and characteristic scale are qualitatively consistent with a positive annihilation-induced contribution of \(\Delta\sim15~{\rm MeV}\). The benchmark counting underestimates the absolute binding energies of both systems by comparable amounts, suggesting that much of the missing attraction is common to the two \(DK\) channels and largely cancels in their difference.

The \(D_{s1}(2460)\), interpreted as an isoscalar \(D^*K\) configuration, provides a closely related example. Replacing \(D\) by \(D^*\) leaves the relevant light-quark topology and benchmark bond counting unchanged, so that the same kaon-assisted covalent and annihilation contributions give \(B_{\rm ann}=\Lambda-2\epsilon+\Delta\sim25~{\rm MeV}\). Its experimental generalized binding energy of approximately \(45~{\rm MeV}\), nearly identical to that of the \(D_{s0}^*(2317)\), suggests that the attraction missing from the benchmark counting is of similar magnitude in the \(DK\) and \(D^*K\) channels. By the same reasoning, an isovector \(D^*K[1(1^+)]\) partner of the \(T_{c\bar s0}^{a}(2327)\) is naturally expected, as also listed in Table~\ref{tab:krelated}. Heavy-quark spin symmetry suggests that this state should have a generalized binding energy of approximately \(36~{\rm MeV}\), corresponding to a mass of around \(2470~{\rm MeV}\), although the present benchmark bond counting gives only \(B_{\rm cov}=\Lambda-2\epsilon\sim10~{\rm MeV}\). 

The recently observed \(B_{s0}^*(5700)\)~\cite{LHCb:2026knl} may provide a bottom counterpart of the \(D_{s0}^*(2317)\)~\cite{Albaladejo:2026yfz,Sanchez-Illana:2026vql,Hao:2026ppm}. Its generalized binding energy relative to the \(BK\) threshold is approximately \(76~{\rm MeV}\), significantly larger than the benchmark estimate \(B_{\rm ann}=\Lambda-2\epsilon+\Delta\sim25~{\rm MeV}\). This discrepancy further illustrates the exploratory nature of the present kaon-related extension. The \(\Lambda(1405)\) and \(\Omega(2012)\) are included in Table~\ref{tab:krelated} only as possible baryonic analogues, for which the nearby compact states relevant to annihilation-induced mixing are not clearly identified. Further studies of kaon-related systems and systems involving other pseudoscalar mesons may help clarify whether the effective kaon-assisted covalent contribution introduced here reflects a more general mechanism associated with pseudoscalar-meson clouds, chiral dynamics, and coupled-channel effects, or instead represents a feature specific to the kaon sector.

\section{Summary}
\label{sec:summary}

In Refs.~\cite{Chen:2021xlu,Chen:2025jgi}, we successively introduced the covalent, creation, and annihilation bonds as effective mechanisms for organizing the residual interactions among color-singlet hadronic clusters. In Ref.~\cite{Chen:2021xlu}, we also constructed a simple toy model for the covalent bond, which is further developed here to incorporate the creation and annihilation bonds within a unified framework. With the generalized binding energy \(B=M_{\rm threshold}-M\), the same framework can be applied to both below-threshold bound states and above-threshold resonances. Throughout this series of studies, we have restricted ourselves to ground-state molecular configurations, while possible excited counterparts, such as the recently observed \(X(3790)^-\)~\cite{BESIII:2026emz}, which was interpreted as a possible $P$-wave $D\bar D$ molecular state in Ref.~\cite{Jiang:2026npd}, are left for future investigation.

For a configuration governed primarily by the covalent-bond mechanism, the generalized binding energy is parameterized as
\begin{equation}
B_{\rm cov}=N_SS+N_WW+N_\Lambda\Lambda-N_RR-N_h\epsilon .
\end{equation}
The benchmark values \(S\sim23~{\rm MeV}\), \(W\sim17~{\rm MeV}\), \(\Lambda\sim18~{\rm MeV}\), \(R\sim14~{\rm MeV}\), and \(\epsilon\sim4~{\rm MeV}\) were estimated phenomenologically from light nuclei and hypernuclei. With these transferable parameters, such configurations generally yield \(B_{\rm cov}>0\) and therefore tend to lie below their corresponding hadronic thresholds.

The creation bond can be incorporated into the same toy model by activating additional sea light-quark degrees of freedom to generate effective covalent correlations when direct valence exchange is unavailable. Its generalized binding energy is written as
\begin{equation}
B_{\rm cre}=\eta_C\left(N_SS+N_WW+N_\Lambda\Lambda\right)-E_{\rm act}-N_RR-N_h\epsilon .
\end{equation}
Here, the additional color-suppression factor is taken to be \(\eta_C=1/9\), while \(E_{\rm act}\) represents the effective excitation energy required for the coherent sea four-quark component to participate in the interhadronic correlation. Taking the \(Z_c(3900)\) as the benchmark gives \(E_{\rm act}\sim17~{\rm MeV}\). Although the activated sea component provides attraction, the resulting gain is color suppressed and generally insufficient to compensate the excitation cost \(E_{\rm act}\). Consequently, creation-bond configurations tend to exhibit \(B_{\rm cre}<0\) and appear above their corresponding hadronic thresholds.

Intriguingly, the extracted value \(E_{\rm act}\sim17~{\rm MeV}\) is numerically close to the reported mass of the \(X_{17}\) particle~\cite{Krasznahorkay:2015iga}:
\begin{equation}
E_{\rm act}
\sim
M_{X17}
=
16.70
\pm 0.35
\pm 0.50
~{\rm MeV}.
\end{equation}
In Ref.~\cite{Chen:2020arr}, we interpreted the \(X_{17}\) particle as a low-energy bare four-quark structure composed of two chiral \(u/d\) quarks and two chiral \(u/d\) antiquarks. The correspondence \(E_{\rm act}\sim M_{X17}\) may therefore suggest that the coherent sea four-quark excitation involved in the creation bond and the bare four-quark configuration proposed for the \(X_{17}\) particle reflect a common low-energy (four-quark excitation) scale in the nonperturbative QCD vacuum. Such a scale may represent an emergent low-energy manifestation of confinement-related dynamics, but this possible correspondence does not imply a microscopic identity of the two four-quark configurations.

The annihilation bond completes the present toy model by allowing a pre-existing covalent- or creation-bond configuration containing an available valence \(q\bar q\) pair to mix with a nearby compact state with compatible quantum numbers through valence \(q\bar q\) annihilation. Its net spectral effect is represented by a signed correction,
\begin{equation}
B_{\rm ann}=B_{\rm cov/cre}+\Delta .
\end{equation}
The \(X(3872)\) provides the benchmark value \(\Delta\sim+15~{\rm MeV}\), corresponding to a downward shift of the predominantly \(D\bar D^*\) eigenstate through mixing with a nearby compact charmonium configuration. When no nearby compact state can generate a well-defined mixing-induced mass shift, annihilation-induced rearrangement may instead contribute to the broadening or dissolution of the original two-hadron component.

Taken together, these three expressions define a compact hierarchy in which the covalent bond provides the basic bond-counting framework, the creation bond supplements it with the additional color suppression and excitation cost associated with the coherent sea four-quark component, and the annihilation bond accounts for the signed spectral shift induced by mixing with a nearby compact state. Several qualitative predictions and extensions follow directly from this unified framework:
\begin{itemize}

\item The absence of identical-core exchange repulsion in the bottom--charm system, together with the expected decrease of the residual kinetic correction for heavier constituents, leads to \(B_{\rm cov}(T_{bc})>B_{\rm cov}(T_{bb})>B_{\rm cov}(T_{cc})\).

\item The \(P_{cs}(4459)\) is interpreted as a bound \(\bar D^*\Xi_c\) covalent configuration, whereas a hypothetical isovector \(D^*\Xi_c\) system is predicted to be a creation-bond configuration with \(B_{\rm cre}\bigl(D^*\Xi_c[I=1]\bigr)\sim-25~{\rm MeV}\), and therefore to lie substantially above threshold.
    
\item Creation--annihilation complementarity suggests characteristic isospin-dependent mass splittings of order \(|\Delta|\sim15~{\rm MeV}\) between otherwise analogous isovector and isoscalar baryon--antibaryon configurations, such as the \(\Xi_c\bar\Xi_c\) systems with \(I=1\) and \(I=0\), provided that a nearby compact state with compatible quantum numbers is available in the annihilation-allowed channel.

\item Motivated by light-nuclear examples, the same bond-counting picture can be extended schematically to multihadron systems, where several active light quarks may occupy a common antisymmetrized multicenter wave function and form connected bond networks. Representative examples are multi-\(\bar D^*\) configurations, the \(\bar D^*\bar D^*\Sigma_c\) system, and the \(D^*\bar D^*B^*\bar B^*\) system.
        
\end{itemize}

The present toy model nevertheless has clear limitations. In particular, several kaon-related systems are not straightforwardly accommodated within the minimal hadronic-bond framework, and therefore motivate an exploratory extension in which the pseudoscalar kaon is treated as an active composite degree of freedom. The hadronic-bond model should therefore be regarded as a simple low-energy organizing principle rather than as a replacement for microscopic QCD or as a tool for precision determinations of pole positions. Its usefulness will ultimately be determined by whether this compact set of phenomenological parameters and bond-counting rules can consistently correlate different near-threshold systems and, more importantly, whether the characteristic predictions identified above are confirmed or falsified by future experiments.

\bibliographystyle{elsarticle-num}
\bibliography{ref}

\end{document}